\documentclass[9pt]{opticajnl}
\journal{opticajournal} 

\setboolean{shortarticle}{true}

\usepackage[group-digits=none]{siunitx}
\graphicspath{{./img/}}
\DeclareMathOperator{\rank}{rank}

\title{High-Resolution Speckle-based Single-Pixel Imaging using Silicon Photonic Non-Redundant Optical Phased Array}
\author[1,*]{Keita Hirashima}
\author[1,2]{Taichiro Fukui}
\author[1,3]{Kento Komatsu}
\author[1]{Chun Ren}
\author[1,4]{Yoshiaki Nakano}
\author[1,$\dagger$]{Takuo Tanemura}
\affil[1]{School of Engineering, The University of Tokyo, 7-3-1 Hongo, Bunkyo, Tokyo 113-8656, Japan}
\affil[2]{Currently with Institute of Electromagnetic Fields (IEF), ETH Zurich, Zurich 8092, Switzerland}
\affil[3]{Currently with Sumitomo Electric Industries, Ltd., 1 Tayacho, Sakae-ku, Yokohama 244-8588, Japan}
\affil[4]{Currently with Toyota Technological Institute, 2-12-1 Hisakata, Tempaku-ku, Nagoya, 468-8511, Japan}
\affil[*]{keita.hirashima@tlab.t.u-tokyo.ac.jp}
\affil[$\dagger$]{takuo.tanemura@tlab.t.u-tokyo.ac.jp}

\begin{abstract}
Optical phased arrays (OPAs) are promising wavefront controlling devices for imaging applications due to their compact and high-speed nature.
However, conventional periodic OPAs have a limited spatial resolution, which scales only linearly with the number of optical antennas $N$.
Here, we experimentally demonstrate high-resolution imaging using a non-redundant OPA (NR-OPA).
Due to the non-redundant antenna layout of NR-OPA based on the Costas array, the resolution scales quadratically with $N$. Combined with the speckle-based single-pixel imaging (SSPI) scheme, which avoids the need for precise phase calibration, we experimentally achieve a large number of resolvable imaging points exceeding 10,000 using a silicon photonic NR-OPA chip with only $N=127$ without sweeping the wavelength.
The demonstrated scheme provides a promising route toward mega-pixel imaging by a compact OPA chip with reduced number of phase shifters.
\end{abstract}

\setboolean{displaycopyright}{false}

\renewcommand*{\journalshorttype}{Preprint}
\dates{}
\doi{}

\begin{document}

\maketitle

\section{Introduction}
An optical phased array (OPA) is a compact and high-speed wavefront generating device that emits phase-controlled light from $N$ antennas.
It can be employed in a wide range of applications,
such as light detection and ranging (LiDAR) \cite{Hsu2021,Poulton2019,Poulton2022,Chung2018,Xu2023,Chen2024},
free-space optical communication \cite{Rabinovich2016,Poulton2019,Kim2024-te}, optical projection \cite{Rekhi2015,Shin2020}, 
photonic switching \cite{Tanemura2011-wh,Kwack2012-bs,Soganci2012-am},
optogenetics \cite{Sacher2022-vh}, optical tweezers \cite{Sneh2024-vr},
and imaging \cite{Komatsu2017-lk,Li2019-tl,Kohno2019,Fukui2021JLT,Wang2021,Hu2023-wk}.
In particular, imaging is a promising application due to the potential for high-speed, compact, and low-cost system offered by silicon photonic integration.
While conventional infrared imaging systems rely on bulky components, which limit their use in spatially constrained environment,
an OPA enables complex wavefront synthesis by a millimeter-scale chip.

One of the challenges in realizing OPA-based imaging systems is the calibration of optical phase deviations induced by the fabrication errors and environmental changes 
\cite{Takahashi2023}.
A promising approach to mitigate this issue is to employ the speckle-based single-pixel imaging (SSPI) \cite{Komatsu2017-lk,Kohno2019,Welsh2013} or computational ghost imaging \cite{Shapiro2008,Wang2021,Hirata2025PhasebiasblindCG} method.
In SSPI, the phase shifters of OPA are driven by pseudo-random signals instead of precisely determined conditions, and the speckle patterns generated from the OPA are used to sample the target information.
Consequently, time-consuming phase calibration procedure can be eliminated \cite{Hirata2025PhasebiasblindCG} or replaced by a substantially simpler step of acquiring generated speckle patterns prior to imaging the target \cite{Kohno2019}.
Moreover, SSPI can mitigate the requirements for phase shifters \cite{Emara2021-ff, Gaolei2025-oc} and is compatible with the compressed sensing schemes to reduce the required number of measurements \cite{Welsh2013,Duarte2008}.

Another challenge of OPA-based imaging system is its spatial resolution.
For the conventional periodic OPAs with antennas located on a uniform grid, the number of resolvable points scales only linearly with the number of antennas $N$.
Simple scaling of $N$ introduces additional issues such as
increased costs in packaging, power consumption, and phase calibration.
One common solution to this problem is sweeping the laser wavelength for 2D beam scanning \cite{Chung2018,Poulton2022}, but it comes at the expense of overall system complexity.
To increase spatial resolution without sweeping the wavelength, it is effective to place optical antennas aperiodically with non-uniform spacing 
\cite{Hulme2015,Hutchison2016,Fatemi2019-yw,Shin2020,Li2021,Huang2023,Qiu2024}.
In particular, the authors have proposed a non-redundant OPA (NR-OPA) \cite{Fukui2021Optica}, which was further investigated by other groups \cite{deCea2024,Midkiff2024}.
In NR-OPA, the antennas are located sparsely so that the autocorrelation function of the antenna layout is widely distributed with minimal overlap. 
Due to this non-redundancy, we have theoretically revealed that significantly higher-resolution SSPI is possible using NR-OPA \cite{Fukui2022}.

In this paper, we employ a silicon photonic NR-OPA chip with only $N=127$ to experimentally demonstrate SSPI with over 10,000 resolvable points.
This is, to our knowledge, the largest number of resolvable points obtained in an integrated-OPA-based imaging system without using the wavelength scanning scheme.

\section{Concept of SSPI using NR-OPA}
\begin{figure}
    \centering
    \includegraphics[width=0.9\linewidth]{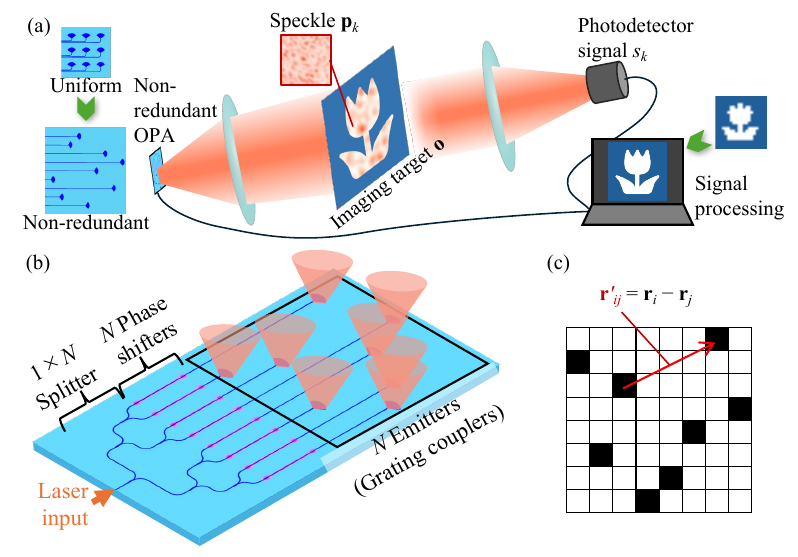}
    \caption{
        (a) Schematic of a speckle-based single pixel imaging system using non-redundant OPA (NR-OPA).
        (b) The schematic of NR-OPA.
        (c) Costas array for $N=8$.
    }
    \label{fig:1}
\end{figure}

Figure~\ref{fig:1}(a) shows the schematic of the SSPI system using OPA.
Inside the OPA chip [\figurename~\ref{fig:1}(b)], the input light is split into $N$ waveguides.
Light passing through each of the waveguides is phase controlled by a phase shifter and is emitted from a grating antenna.
By driving $N$ phase shifters of the OPA using random electrical signals, speckle-like far field patterns (FFPs) are generated by interference of the light from $N$ emitters on the OPA.
The generated intensity patterns are used to illuminate the imaging target, and the total optical power transmitted from the target is detected by a single-pixel photodetector (PD).
To retrieve the image, the illumination patterns are switched for $K$ times, and $K$ PD signals are recorded.

The $k$-th PD signal $s_k$ can be expressed using the transmittance distribution of the target object $\mathbf{o}$ and the $k$-th illumination pattern $\mathbf{p}_k$ as $s_k = \mathbf{p}_k \cdot \mathbf{o}$.
Thus, by switching the illumination patterns $K$ times, we have a simultaneous linear equation
\begin{equation}
    \mathbf{s}=\mathbf{P}\mathbf{o},
    \label{eq:SPO}
\end{equation}
where $\mathbf{s} = \left(s_1\ s_2\ \cdots\ s_K\right)^\top$ and $\mathbf{P} = \left(\mathbf{p}_1\ \mathbf{p}_2\ \cdots\ \mathbf{p}_K\right)^\top$.
We can reconstruct the target image $\mathbf{o}$ by calculating
\begin{equation}
    \mathbf{o}=\mathbf{P}^+ \mathbf{s} ,
    \label{eq:OPS}
\end{equation}
where $\mathbf{P}^+$ is the Moore--Penrose pseudoinverse of $\mathbf{P}$ \cite{Kohno2019}.
Through the singular value decomposition (SVD) described as $\mathbf{P} = \mathbf{U}\mathbf{\Sigma}\mathbf{V}^\dagger$,
it can be written as $\mathbf{P}^+ = \mathbf{V}\mathbf{\Sigma}^+\mathbf{U}^\dagger$, where $\mathbf{\Sigma}^+$ is formed by taking the reciprocal of each non-zero singular value on the diagonal and transposing the resulting matrix.

In the NR-OPA, the antennas are located based on the concept of NRA, where the displacements between every pair of elements ($\mathbf{r}^\prime_{ij}$ in \figurename~\ref{fig:1}(c)) are mutually distinct. 
In this work, we employ a silicon photonic NR-OPA based on the Costas array \cite{Fukui2021Optica}, which is a 2D NRA formed by selecting $N$ points on a 2D $N \times N$ grid.
An example of Costas array ($N=8$) is displayed in \figurename~\ref{fig:1}(c).
Theoretically, the number of resolvable points 
can be evaluated by 
$\rank(\mathbf{P})$ \cite{Fukui2021-te}, which indicates the amount of independent information obtained about the object $\mathbf{o}$ from Eq.~(\ref{eq:SPO}), and is equal to the number of distinct displacement vectors between emitters. 
For the case of NR-OPA, since these displacement vectors are mutually distinct, the imaging resolution is maximized to $N(N-1) + 1 = N^2 - N + 1$.
\cite{Fukui2022}

\section{Experimental setup}
\begin{figure}[h]
    \centering
    \includegraphics[width=0.75\linewidth]{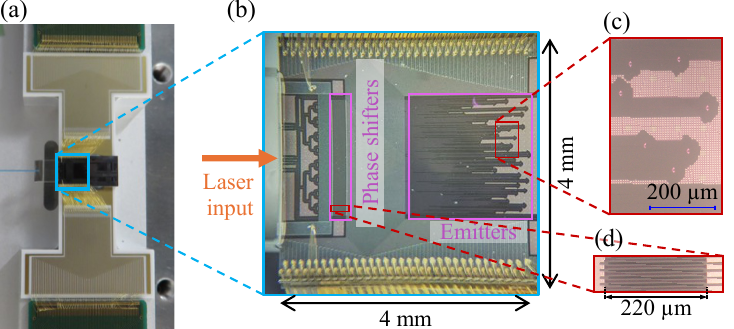}
    \caption{Silicon photonic NR-OPA chip with $N=127$.
    (a) Mounted chip with an PM fiber and all 127 phase shifters electrically connected to the driver circuit. 
    (b) A micrograph of the NR-OPA chip.
    (c, d) Magnified micrographs of (c) grating couplers arranged on a Costas array and (d) thermo-optic phase shifters.
    }
    \label{fig:chip}
\end{figure}

Figure~\ref{fig:chip} shows the silicon photonic NR-OPA chip with $N=127$ employed in this work \cite{Fukui2021Optica}, which was fabricated by a multi-project wafer foundry service.
A silicon-on-insulator wafer with 220-\si{nm}-thick Si layer and 2-\si{\um}-thick buried oxide layer was used.
The OPA consisted of an edge coupler for a fiber input, a cascade of 1$\times$2 multi-mode interference couplers (MMIs) to split the light to 128 waveguides (with one port used for monitoring), 127 thermo-optic phase shifters (\figurename~\ref{fig:chip}(d)), and 127 grating couplers placed on a Costas array (\figurename~\ref{fig:chip}(c)).
The grid pitch of the Costas array was \SI{15}{\um}, resulting in a field of view (FOV) of \SI{5.92}{\degree} at a wavelength of \SI{1550}{\nm}.
Each grating coupler was fan-shaped, with a length of \SI{10}{\um} and a width of \SI{16}{\um}. The waveguides connecting to the grating couplers were routed to have identical lengths.
As a thermo-optic phase shifter, we employed \num{220}-\si{\um}-long TiN heater with \num{1}-\si{\kilo\ohm} electrical resistance and \num{4}-\si{\um}-wide trenches for thermal isolation, achieving $2\pi$ phase shift at approximately \SI{20}{\mW}.
All heater electrodes were connected to a driver circuit via wire bonding.

\begin{figure}[h]
    \centering
    \includegraphics[width=0.75\linewidth]{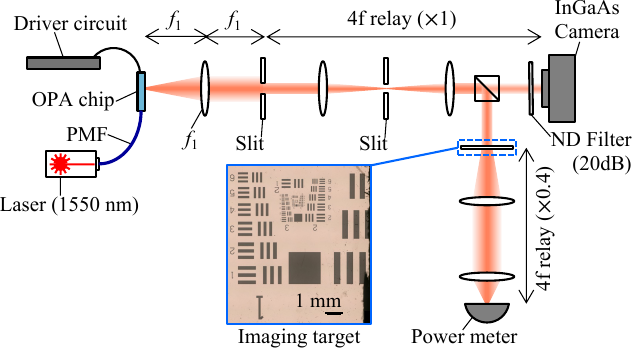}
    \caption{Experimental setup. The FFP of the OPA chip is cropped and relayed to the imaging target. The transmitted light from the target is collected by a power meter. The inset shows a microscopic image of the target. $f_1$=\SI{100}{\mm}.}
    \label{fig:setup}
\end{figure}

Figure~\ref{fig:setup} shows the experimental setup.
Continuous-wave 1550-\si{\nm}-wavelength light was input to the NR-OPA chip through a polarization-maintaining fiber.
The FFP of the OPA obtained through a lens ($f_1=\SI{100}{mm}$) was cropped using the first slit to the imaging region smaller than the FOV of the OPA. 
It was then relayed to the USAF 1951 resolution test chart (\figurename~\ref{fig:setup} inset) through a 4f imaging system. The second slit was placed to spatially filter out stray light scattered at the input facet of the OPA chip.
The transmitted light from the imaging target was collected and the total power $\mathbf{s}$ was measured by an optical power meter (HP 81524A).
The 127 phase shifters on the OPA were driven by pseudo-random voltage signals in a range from 2 V to 5 V, such that the phases, which are proportional to the square of the voltages, were uniformly distributed.
For convenience, an InGaAs camera (Hamamatsu Photonics C12741-03) with a 20-dB neutral-density (ND) filter was used to capture the actual patterns $\mathbf{P}$, illuminated to the sample.
From a separate measurement without the cropping slits, the FOV was measured to be \SI{10.56}{\mm} at the plane of the imaging target, which was close to the designed value.

From the measured $\mathbf{s}$ and $\mathbf{P}$, the image was retrieved through Eq.~(\ref{eq:OPS}). To mitigate memory consumption, the pseudoinverse was computed by applying a QR decomposition and subsequently performing SVD on the resulting upper triangular matrix $R$.
Additionally, to avoid amplifying the noise associated with small singular values (SVs) in $\mathbf{\Sigma}$, we employed the truncated SVD method \cite{Hansen1987}; the small SVs beyond a cutoff index were set to zero.

\section{Results and Discussion}
\begin{figure}[h]
    \centering
    \includegraphics[width=0.75\linewidth]{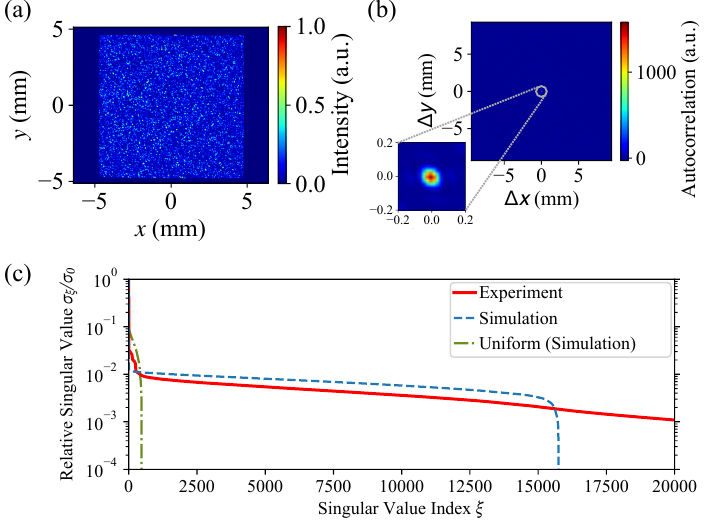}
    \caption{
        (a) An example of captured illumination patterns.
        (b) 2D autocorrelation of the illumination pattern in (a), exhibiting a sharp peak at the origin and near-zero values elsewhere. The inset shows a magnified view of the central peak.
        (c) Top 20,000 singular values of the illumination matrix.
    }
    \label{fig:result-illum}
\end{figure}

Figure~\ref{fig:result-illum}(a) shows an example of observed intensity pattern, illuminated to the target.
To examine the randomness of the illumination pattern,
its 2D autocorrelation is plotted in \figurename~\ref{fig:result-illum}(b).
The autocorrelation has a sharp peak at the center with full width at half maximum of \SI{0.07}{\mm} (equivalent to \SI{0.04}{\degree}), which represents the fine-grained nature of the speckle.
Additionally, the autocorrelation values remain close to zero in all other regions, implying the overall spatial randomness of the pattern and ensuring that any grating lobes or pattern repetitions fall outside the crop region.

To evaluate the amount of spatial information that can be extracted by 65,536 randomly generated illumination patterns ($K=65{,}536$), the SVs of the illumination matrix $\mathbf{P}$ are calculated.
Figure~\ref{fig:result-illum}(c) shows the SVs for the experimentally obtained illumination matrix (red-solid),
along with numerically simulated results for the NR-OPA (blue-dashed) and for a uniform OPA (green-dash-dotted).
The SVs are sorted in descending order and numbered using an index $\xi$. They are then normalized by the largest SV.
Unlike the uniform OPA, whose SV drop rapidly at $\xi \sim 500$, the experimentally obtained SVs for the NR-OPA remain relatively large up to $\xi = 16,003\ (= N^2-N+1)$, in good agreement with the simulated result.

\begin{figure}[t]
    \centering
    \includegraphics[width=0.6\linewidth]{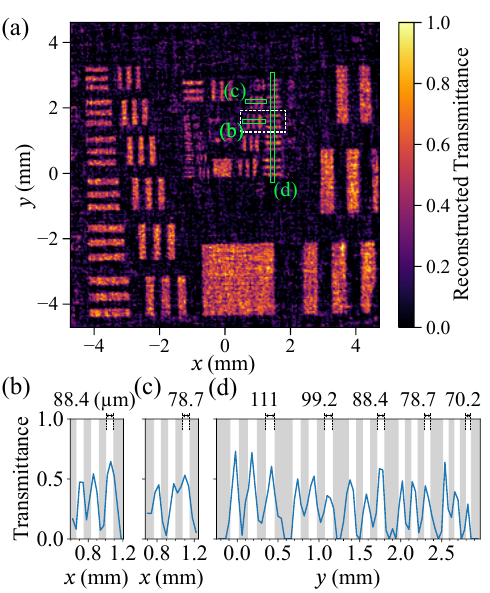}
    \caption{
        (a) A reconstructed image of the target with \SI{9.44}{\mm}$\times$\SI{9.32}{\mm} region and an image size of 236$\times$\SI{233}{px} ($K=65,536$, $\xi_{\max}=18000$). The white dashed rectangle shows the finest patterns resolved by our device.
        (b,c) Horizontal and (d) vertical cross sections at the regions outlined by green lines in (a). The gray shaded regions represent the pattern of the imaging target.
    }
    \label{fig:result-reconst}
\end{figure}

Figure~\ref{fig:result-reconst}(a) shows a reconstructed \SI{9.44}{\mm}$\times$\SI{9.32}{\mm} image with 40-\si{\um} pixel size, obtained using the illumination patterns presented in \figurename~\ref{fig:result-illum}.
The cutoff index of the truncated SVD was set to $\xi_{\max} = 18,000$.
The peak signal-to-noise ratio (PSNR) is \SI{13.9}{\dB}, which could be further improved by increasing the optical signal-to-noise ratio of the measurement and optimizing the illumination patterns \cite{Emara2021-an}.

To examine the spatial resolution in detail, cross-sectional plots around the smallest resolved element on the USAF 1951 chart along horizontal and vertical directions are presented in \figurename~\ref{fig:result-reconst}(b-d).
We can confirm from \figurename~\ref{fig:result-reconst}(b) and \figurename~\ref{fig:result-reconst}(d) that line and space patterns with widths as narrow as \SI{88.4}{\um} are successfully resolved, whereas the image starts to blur for narrower patterns as shown in \figurename~\ref{fig:result-reconst}(c).
From this result, we derive the spatial resolution to be \SI{5.66}{lp/\mm}, or equivalently, \SI{11.32}{points/\mm}.
Since the total imaging area is \SI{9.44}{\mm}$\times$\SI{9.32}{\mm}, the total number of resolvable points is as large as ($107\times106=$) \num{1.13e4}.

\section{Conclusion}
We have experimentally demonstrated high-resolution speckle-based OPA imaging at a single wavelength.
Using a compact silicon photonic NR-OPA chip with only 127 phase shifters ($N=127$), we have successfully retrieved a 2D image with \num{1.13e4} resolvable points, which is, to our knowledge, the highest spatial resolution obtained by OPA-based imaging systems without sweeping the wavelength.
This is owing to the non-redundant antenna arrangement of our NR-OPA based on the Costas array, which enables the spatial resolution to scale quadratically with the number of antennas $N$. Such feature has distinct advantage over the conventional OPA with uniform spacings, where the resolution scales only linearly with $N$.
By using a larger-scale NR-OPA with $N \sim 10^3$, which is feasible with the state-of-the-art OPA technologies \cite{Chung2018, Poulton2022}, megapixel-order imaging should be realized.
The demonstrated scheme should thus pave the way towards the realization of high-resolution low-cost imaging systems using OPAs.

\begin{backmatter}
\bmsection{Funding}
Japan Society for the Promotion of Science (JSPS) KAKENHI (JP23H05444);
Japan Science and Technology Agency (JST) SPRING (JPMJSP2108) (K.H.)

\bmsection{Acknowledgement}
Portions of this work were presented at the Conference on Lasers and Electro-Optics (CLEO) in 2025, SS174\_3.
Part of the results in this research were obtained using supercomputing resources at Cyberscience Center, Tohoku University.

\bmsection{Disclosures}
The authors declare no conflicts of interest.

\bmsection{Data Availability}
Data underlying the results presented in this paper are not publicly available at this time but may be obtained from the authors upon reasonable request.
\end{backmatter}

\bibliography{references}

\end{document}